# A photonic thermalization gap in disordered lattices

H. Esat Kondakci, Ayman F. Abouraddy and Bahaa E. A. Saleh

CREOL, The College of Optics & Photonics, University of Central Florida, Orlando, Florida 32816, USA

The formation of gaps – forbidden ranges in the values of a physical parameter – is a ubiquitous feature of a variety of physical systems: from energy bandgaps of electrons in periodic lattices[1] and their analogs in photonic[2], phononic[3], and plasmonic[4] systems to pseudo energy gaps in aperiodic quasicrystals.[5] Here, we report on a 'thermalization' gap for light propagating in finite disordered structures characterized by disorder-immune chiral symmetry[6] – the appearance of the eigenvalues and eigenvectors in skew-symmetric pairs. In this class of systems, the span of sub-thermal photon statistics is inaccessible to input coherent light, which – once the steady state is reached – always emerges with super-thermal statistics no matter how small the disorder level. We formulate an independent constraint that must be satisfied by the input field for the chiral symmetry to be 'activated' and the gap to be observed. This unique feature enables a new form of photon-statistics interferometry: the deterministic tuning of photon statistics – from sub-thermal to super-thermal – in a compact device, without changing the disorder level, via controlled excitation-symmetry-breaking realized by sculpting the amplitude or phase of the input coherent field.

For electrons in crystals, lattice symmetries play a critical role in establishing energy gaps, and introducing disorder typically diminishes their role [1]. One exception lies in certain disorder-immune symmetries that emerge in random matrix theory [7], such as chiral [6] and particle-hole symmetric ensembles [8], which play decisive roles in diverse areas of physics ranging from superconductivity [9] to quantum chromodynamics [10]. A hallmark of disorder-immune chiral symmetry [7,11] is that the system Hamiltonian can be transformed into a block off-diagonal matrix representation, corresponding to separate bipartite sublattices [12]. We elucidate this concept in the context of the one-dimensional (1D) disordered tight-binding lattice models depicted in Fig. 1a,b, one of which maintains chiral symmetry. In Fig. 1a, coupling between lattice sites is fixed $C_{x,x+1} = \bar{C}$ ($x$ is the site index) while their energies $\beta_x$ are randomly perturbed – so-called diagonal disorder [13]. Alternatively, these energies may be held fixed $\beta_x = \bar{\beta}$, while the couplings are perturbed – so called off-diagonal disorder (Fig. 1b) [14]. Only in the latter case can the Hamiltonian be cast in block off-diagonal form by dividing the Hilbert space into subspaces of even- and odd-indexed lattice sites. Such lattices provide a setting for studying disorder-immune chiral symmetry, which is also realized in two-dimensional (2D) lattices such as square, hexagonal [11,12], and even-sited ring lattices and certain Penrose tilings [15] under conditions of off-diagonal disorder. A Hamiltonian endowed with chiral symmetry features eigenvalues and eigenvectors that occur in skew-symmetric pairs in every realization of the disorder. Chiral ensembles raise fundamental questions regarding the impact on transport of the interplay between disorder and symmetry. Specifically, can one detect unambiguous traces of the underlying symmetry in the statistics of the transported wave even at maximal lattice disorder?

In investigating these questions, optics provides a particularly useful platform to explore random matrix theory and the ramifications of disorder-immune symmetries. Since the randomness of the lattice influences the propagating light wave, the emerging random light must be described using the tools of statistical optics and optical coherence theory. Starting from the Helmholtz wave equation for monochromatic light, we obtain a Schrödinger-like equation in the paraxial limit [16,17],

$$i\frac{\partial A}{\partial z} + \frac{1}{2k_o n_o}\nabla_T^2 A + k_o \Delta n(x,y) A = 0; \qquad (1)$$

where the axial propagation coordinate $z$ plays the role of time, $\nabla_T$ is the transverse Laplacian along the $x$ and $y$ coordinates, $A(x,y,z)$ is the slowly varying optical field envelope, $k_o$ is the free-space wave number, and $\Delta n(x,y)$ is an axially invariant perturbation to an averaged refractive index $n_o$ [17]. If the index features continuous but localized perturbations forming a disordered lattice, then the system is akin to a set of parallel coupled waveguides. Assuming the perturbations allow for single-mode waveguides with nearest-neighbor-only coupling, then the above equation can be mapped onto a set of coupled equations for the complex amplitudes $E_x$ at the lattice sites (see Supplementary) [18],

$$-i\frac{dE_x}{dz} = \beta_x E_x + C_{x,x+1} E_{x+1} + C_{x,x-1} E_{x-1}. \qquad (2)$$

In a finite-width lattice of $2N+1$ waveguides, labeled $x = -N, \ldots, N$, $\beta_x$ is the propagation constant of the $x^{\text{th}}$ waveguide mode and $C_{x,x+1}$ is the coupling coefficient of waveguides $x$ and $x+1$. This "discretized" model (equation (2)) of the paraxial Helmholtz equation (equation (1)) allows us to obtain, in a photonics setting, tight-binding Hamiltonian models with precise control over their disorder – leading to optical realizations of Anderson localization [17-20] and topological insulators [21], among a wealth of examples [22-24]. We adopt here these arrays of coupled optical waveguides as the realization of choice (Fig. 1c) [19] – but our results are applicable to other embodiments [17, 25-27].

One intuitively expects that monochromatic light propagating along a transversely disordered lattice of the kind described above will gradually become spatially randomized [28]. Here, the randomness is not manifested as time fluctuations, but rather as uncertainty defined in the probability space of disorder realizations. This model is commonly used in optical coherence theory [29] to describe light propagation through a random inhomogeneous medium or scattering from a rough surface and forming a speckle pattern. The propagating light can acquire thermal statistics much like thermal light emitted from a blackbody radiator or resulting from the interference of many independent random electromagnetic wave packets. Since it remains monochromatic and does not exhibit time fluctuations, it is often referred to as pseudo-thermal light. Would the light transmitted in a disordered lattice ultimately 'thermalize,' regardless of any inherent disorder-immune symmetry? In this paper, we unveil a surprising optical phenomenon manifested in the intensity statistics, and the associated photon statistics, when the disordered lattice has underlying chiral symmetry. Instead of the expected gradual loss of coherence with increasing disorder, detailed numerical calculations and analytical modeling reveal that an abrupt transition occurs in the intensity statistics – from coherent light to super-thermal light – for vanishingly small values of off-diagonal disorder when the steady state is reached. In other words, there is a 'thermalization gap' – the sub-thermal range is inaccessible regardless of the disorder level. This gap does *not* exist in systems lacking chiral symmetry (e.g., diagonal disorder). These phenomena are robust and they emerge in *finite* lattices with even a few sites. Furthermore, chiral symmetry is a necessary but not sufficient condition for the emergence of a photonic thermalization gap: the input field distribution itself must satisfy an independent constraint. We exploit this feature to devise a methodology for 'photon-statistics interferometry': deterministic tuning of the intensity and photon statistics in a compact device by simply controlling the amplitude or phase of a coherent input field.

We start by evaluating the statistics of the intensity $I_x(z)$ at site $x$ and axial position $z$ when a single site $x = 0$ is excited at $z = 0$ by a monochromatic coherent field (Fig. 1c). Even though the transmitted light remains monochromatic and its intensity undergoes no temporal fluctuations as it travels through the disordered lattice, its coherence properties, defined in terms of ensemble averages, are significantly altered. We choose the normalized mean-square intensity $g_x^{(2)}(z) = \langle I_x^2(z) \rangle / \langle I_x(z) \rangle^2$ as a measure of intensity uncertainty, where $\langle . \rangle$ denotes ensemble averaging [27]. Randomness of the associated photon number $n_{\text{ph}}$ is characterized by the Fano factor $F =$

$\mathrm{Var}(n_\mathrm{ph})/\langle n_\mathrm{ph}\rangle = 1 + \langle n_\mathrm{ph}\rangle(g^{(2)} - 1)$, which is also a measure of photon-bunching [30], the tendency of photons to arrive together and become more correlated when $g^{(2)}$ increases. For coherent light $g^{(2)} = 1$ and $F = 1$, while for thermal light $g^{(2)} = 2$ and $F = 1 + \langle n_\mathrm{ph}\rangle$. By convention, the mutually exclusive spans of $1 < g^{(2)} < 2$ and $2 < g^{(2)}$ are associated with sub-thermal and super-thermal light, respectively [29]. Here, the terms sub-thermal and super-thermal refer to levels of intensity and photon-number uncertainties compared to those characterizing thermal light. The disorder level in a lattice with off-diagonal (diagonal) disorder is quantified by the half-width $\Delta C$ ($\Delta \beta$) of the probability distribution around the mean $\bar{C}$ ($\bar{\beta}$) of the random coefficients $C_{x,x+1}$ ($\beta_x$), and we assume that the probability distribution is uniform. We use dimensionless variables by writing the coupling coefficients and propagation constants in units of $\bar{C}$ and the propagation distance $z$ in units of the coupling length $1/\bar{C}$.

It is important to note that no qualitative features of the averaged intensity $\langle I_x(z)\rangle$ depend on chiral symmetry in transversely disordered lattices [20,31,28] (see Supplementary Fig. S1). However, the higher-order statistics captured in $g_0^{(2)}(z)$ at the central waveguide $x = 0$ reveals a clear demarcation – in its dependence on both $z$ and disorder level – between diagonal and off-diagonal disorder. Using two color palettes to distinguish the domains of sub-thermal and super-thermal statistics, Fig. 2a shows that the optical wave generally evolves to super-thermal light for off-diagonal disorder, and to sub-thermal light for diagonal disorder.

In determining the asymptotic behavior of $g_0^{(2)}(z)$ at large $z$, or $g_0^{(2)}(\infty)$, we first set a criterion for reaching the steady state, whereupon the mean intensity becomes stationary. If the eigenvalues and eigenmodes (or, simply, modes) of the lattice Hamiltonian are $\{b_n\}$ and $\{\phi_n(x)\}$, we take the steady state to be the distance at which $\sigma_n z \gg 2\pi$ for all $n$, where $\sigma_n$ is the standard deviation of the eigenvalue $b_n$. Using this criterion, $g_0^{(2)}(\infty)$ undergoes an abrupt jump from unity to 3 at low off-diagonal disorder levels, followed by a gradual drop to 2 with increasing disorder level $\Delta C$ (Fig. 2b); the thermalization gap $1 < g_0^{(2)}(\infty) < 2$ is evident. That is, the transmitted light is either coherent, under conditions of perfect lattice periodicity, or super-thermal at any level of disorder. For diagonal disorder, on the other hand, the jump occurs from unity to 2 followed by a monotonic decay back to unity with increasing disorder level $\Delta \beta$, and no gap appears.

The statistical measure $g_0^{(2)}$ is a single scalar descriptor of the intensity statistics and – despite its utility – does not constitute a complete representation of the field. A unique description requires knowledge of the full photon-number distribution $P(n_\mathrm{ph})$, which we have plotted in Fig. 2c for selected disorder levels in the asymptotic regime. A sudden transition takes place between the Poissonian statistics in a periodic lattice to super-thermal (sub-thermal) statistics in lattices with (without) chiral symmetry. Diagonal disorder witnesses a transition to an exponential probability distribution of intensity $P(I) = \frac{1}{\mu} e^{-\frac{I}{\mu}}$ associated with Bose-Einstein statistics of the photon number $P(n_{\rm ph}) = \frac{\mu^{n_{\rm ph}}}{(1+\mu)^{n_{\rm ph}+1}}$, where $\mu = \langle n_{\rm ph}\rangle$ is the average photon number. On the other hand, off-diagonal disorder engenders a Gaussian-square probability distribution of intensity $P(I) = \frac{1}{\sqrt{\pi\mu I}} e^{-\frac{I}{\mu}}$ and a modified Bose-Einstein photon statistics $P(n_{\rm ph}) = \frac{(2n_{\rm ph}-1)!!}{n_{\rm ph}! 2^{n_{\rm ph}}} \frac{\mu^{n_{\rm ph}}}{(1+\mu)^{n_{\rm ph}+1/2}}$ [30].

The dramatic distinction between off-diagonally and diagonally disordered lattices has its origin in the disorder-immune chiral symmetry underlying the former – as revealed in their eigenvalues and modes. Chiral symmetry in the system at hand results in mode-pairing such that $b_n = -b_{-n}$ and $\phi_n(x) = (-1)^x \phi_{-n}(x)$ for all $n$ and $x$ [20]. This skew-symmetry in mode pairs extends to all disorder levels and is valid for *each* individual realization from a statistical ensemble. There is no such constraint on lattices with diagonal disorder, where the eigenvalue spectrum has chiral symmetry *only on average*, $\langle b_n \rangle = -\langle b_{-n}\rangle$. To place our findings on a secure foundation, we have established a theoretical model that utilizes this distinction in the limiting cases of low and high disorder. In the former, we make use of perturbation theory, while in the latter we exploit the exponential localization of the random lattice modes. In both cases, the deterministic symmetry that distinguishes chiral ensembles contributes to an increase in field correlations and hence higher $g_0^{(2)}$, while lower-order statistics quantified by the average intensity are impervious to the influence of this disorder-immune symmetry. Additionally, our analytical model reveals the dependence of $g_0^{(2)}$ on the lattice size. The details are involved and are provided in the Supplementary Information. Here we give the analytical formulae for the asymptotic values of $g_0^{(2)}$ as a function of the lattice size $2N + 1$ in the low-disorder limit,

diagonal : $g_0^{(2)}(\infty) = 2 - \frac{1}{N+1}$, (3a)

off-diagonal : $g_0^{(2)}(\infty) = 3 - \frac{3}{N+1} + \frac{p}{(N+1)^2}$; (3b)

and in the high-disorder limit,

diagonal $\Delta\beta \approx 4$ : $g_0^{(2)}(\infty) = 1 + \frac{1}{3}\frac{2N}{2N+1}$, (4a)

off-diagonal $\Delta C \approx 1$ : $g_0^{(2)}(\infty) = 2 - \frac{1}{2N+1} + \frac{2^4 p}{(2N+1)^4}$. (4b)

Here $p = 0$ ($p = 1$) when $N$ is odd (even). These analytic formulae are compared to simulated ensemble averages in Fig. 3 with excellent agreement between the two. We find that the thermalization gap emerges even in small-sized lattices with $\approx 15$ sites.

The results presented thus far have assumed that only a single site is excited, which is a highly symmetric configuration, and it is not clear *a priori* whether the concept of a thermalization gap extends to an arbitrary excitation. We have found that only certain classes of field excitations 'unlock' or 'activate' the underlying chiral symmetry of the Hamiltonian, thereby enabling access to super-thermal statistics. A necessary condition to activate the chiral symmetry is for the mode pairs with indices $\pm n$ to be excited with equal strength. If we write the input field $E_x(z =$

0) in terms of the lattice modes $E_x(0) = \sum_{n=-N}^{N} c_n \phi_n(x)$, where $c_n$ is the amplitude of the $n^{\text{th}}$ mode, then a single-site excitation $E_x(0) = \delta_{x,0}$ results in $c_n = \phi_n(0)$ (all the modes $\phi_n(x)$ and eigenvalues $b_n$ are real). In this case, $|c_n| = |c_{-n}|$ and the chiral symmetry is activated. A sufficient condition for an input field to activate chiral symmetry is that the amplitudes at neighboring sites differ in phase by $\pm \frac{\pi}{2}$ (see Supplementary Information). Guided by this principle, we can devise field distributions that activate or break the chiral symmetry of the system's Hamiltonian, thereby determining the accessible spans of photon statistics at the output of a single system.

In the first approach, we consider a two-site excitation with amplitudes $E_0$ and $E_1$ at sites 0 and 1, respectively (Fig. 4a). Varying the relative phase $\theta$ for equal-amplitude excitation, $E_1 = e^{i\theta} E_0$ results in $c_{\pm n} \propto \phi_n(0) \pm e^{i\theta} \phi_n(1)$, and the requirement $|c_n| = |c_{-n}|$ is satisfied only when $\theta = \pm \frac{\pi}{2}$. At these values $g_0^{(2)}$ attains a maximal value, while minima are reached when $\theta = 0$ or $\pi$ corresponding to maximal chiral-symmetry-breaking. In a compact device, a lattice with 51 sites and length $z = 10$, one may tune $g_0^{(2)}$ continuously between 1.6 (sub-thermal) through 2.5 (super-thermal) for an off-diagonal disorder level of $\Delta C = 0.5$ (Fig. 4c). Alternatively, by varying the relative amplitude $\eta = E_1/E_0$ of the two-site excitation with no relative phase, the modal weights $c_{\pm n} \propto \phi_n(0) \pm \eta \phi_n(1)$ imply gradual chiral-symmetry-breaking from $\eta = 0$ to 1.

In the second approach, the whole array is illuminated and the lattice sites are divided into even- and odd-indexed subsets (corresponding to bipartite lattices [7,11,12]). The excitation amplitudes within each subset, $E_{\text{even}}$ and $E_{\text{odd}}$, are all equal in amplitude and in-phase. We consider the impact of varying the relative phase between $E_{\text{even}}$ and $E_{\text{odd}}$ when they are equal in amplitude (Fig. 4e) and varying their relative amplitude when they are in phase (Fig. 4f). The results are similar to those obtained for the two-site excitation scheme (Fig. 4c,d), except that the tuning of the photon statistics applies at the output to all the lattice sites (except near the edges). Figure 4g-j depicts photon number distributions for $\Delta C = 0.5$ with a mean photon number of 10 corresponding to Fig. 4c-f, respectively. This configuration highlights several critical features of our work. The tuning of the statistics is not a consequence of varying the average intensity, which is fixed, nor the lattice disorder level, which is also fixed. Instead, the underlying chiral symmetry is either activated, thereby granting access to super-thermal statistics, or is gradually de-activated, dropping to sub-thermal statistics.

Our findings reveal profound consequences of the interplay between disorder and symmetry in lattices that are revealed when high-order statistics are probed, but hidden in the usual averaged intensity. Disorder-immune symmetries in particular may impact photon statistics in counter-intuitive ways, as we have shown here in the class of 1D chiral ensembles. Experimental realization of these predictions is within the reach of current photonic fabrication capabilities whether in the context of coupled waveguides [17-20,27], resonators [26], photonic crystals [25], or photorefractive arrangements [17]. Fascinating questions related to entropy generation in the field, the potential impact of nonlinearities induced in the lattice at high fluence levels, the effect of time-varying potentials, and the evolution of non-classical light such as spatially entangled biphoton and Fock states in activated chiral lattices can now be pursued. The question of the existence of disorder-immune symmetries and the associated thermalization gap in quasi-crystals [5] and incommensurate Aubry-André lattice models [32,33] is intriguing. Finally, while we have couched our results in an optical setting, they may be readily mapped onto other physical systems by virtue of the generic tight-binding model we have adopted.

## Acknowledgements

The authors thank the Advanced Research Computing Center at the University of Central Florida for access to the high performance computing cluster.

## Additional information

Supplementary information is available in the online version of the paper.

# Figures

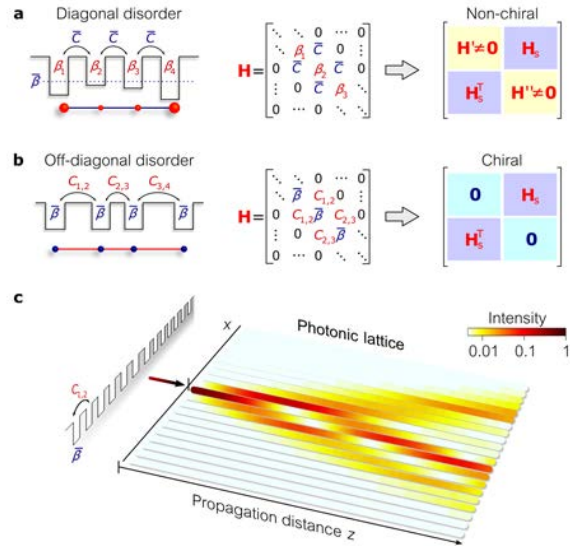

**Figure 1 | One-dimensional lattices with diagonal and off-diagonal disorder.**
**a**,**b**, Coupled potential wells representing one-dimensional lattices characterized by (**a**) diagonal and (**b**) off-diagonal disorder, their corresponding Hamiltonians, and the associated rearranged block off-diagonal matrix forms. We assume nearest-neighbor coupling only. **c**, Schematic representation of a 1D random lattice of coupled optical waveguides excited from a single lattice site, corresponding to a tight-binding model with off-diagonal disorder. Color represents the intensity for a single realization of disorder.

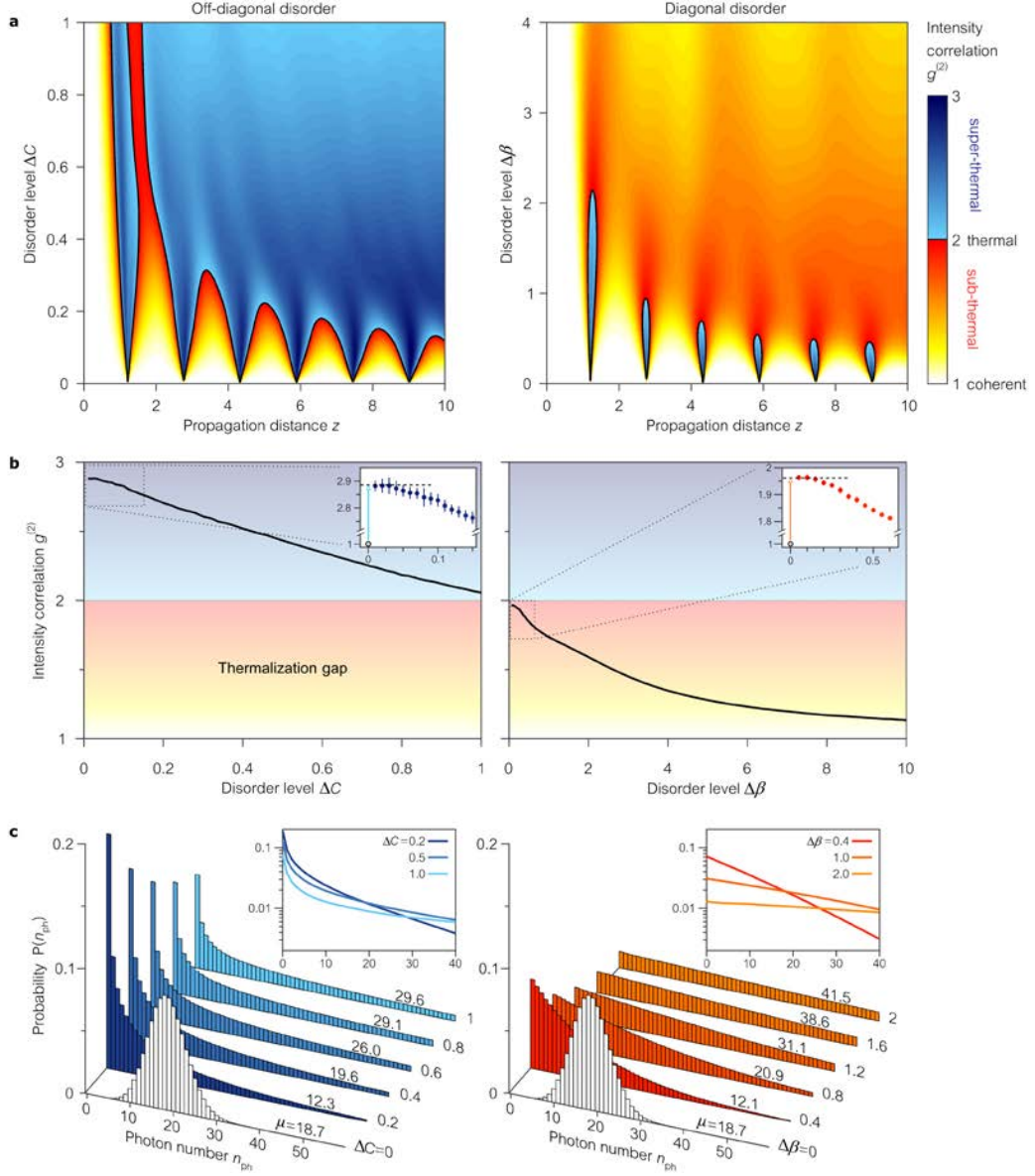

**Figure 2 | Emergence of a thermalization gap in disordered lattices.**
**a**, Normalized intensity correlation $g_0^{(2)}(z)$ at the excitation site ($x = 0$) as a function of disorder level and propagation distance $z$ in the case of (left) off-diagonal and (right) diagonal disorder for lattice size $2N + 1 = 51$. Off-diagonal disorder mainly exhibits super-thermal statistics (blue-black color scheme) whereas diagonal disorder exhibits sub-thermal statistics (white-red color scheme). **b**, Asymptotic $g_0^{(2)}(\infty)$ as a function of the disorder level in the case of (left) off-diagonal and (right) diagonal disordered lattices for $2N + 1 = 51$ and $z = 10^4$ (satisfying the steady-state criterion). A thermalization gap emerges between $g_0^{(2)}(\infty) = 1$ and 2 in off-diagonal disordered lattices. Insets highlight the abrupt jump in $g_0^{(2)}(\infty)$, the dashed lines are obtained from a theoretical model in the low-disorder limit (see text for details), and error bars denote the standard deviation in $g_0^{(2)}$ within the vicinity of $z = 10^4$. **c,** Photon number distributions $P(n_{\text{ph}})$ corresponding to **b** for a coherent input with a fixed mean photon number $\mu \approx 680$ at the input. Insets show $P(n_{\text{ph}})$ for selected disorder levels on a semi-log scale. **a-c**, Ensemble size is $10^6$.

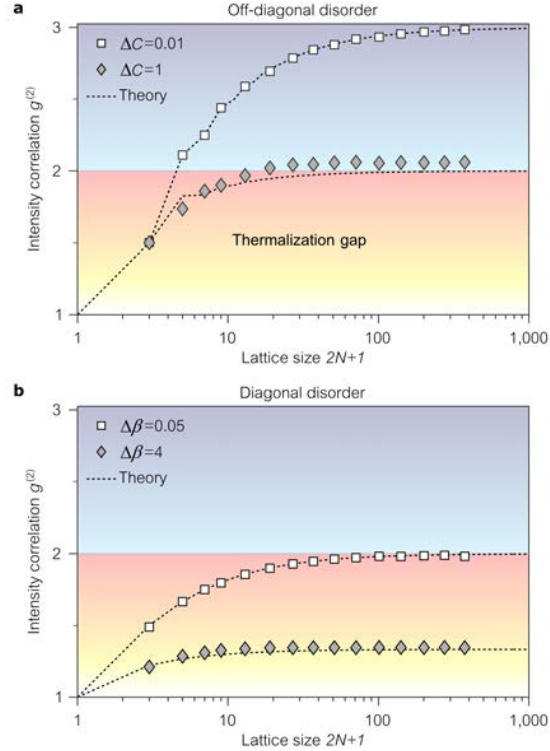

**Figure 3 | Asymptotic normalized intensity correlation as a function of lattice size.**
**a**, Plots of the normalized intensity correlation $g_0^{(2)}(\infty)$ in the steady state in off-diagonal disordered lattices. Squares (diamonds) correspond to a low-disorder (high-disorder) level. Dotted lines are based on an analytical model (see Supplementary Information) given in equations (3) and (4). The background color is associated with the palette used in Fig. 2a and aim at highlighting the regimes of super-thermal (top-half, blue palette) and sub-thermal (bottom-half, red palette) statistics. **b**, Same as (**a**), except that lattices with diagonal disorder are considered. **a**,**b**, Ensemble size is $10^6$ and the asymptotic propagation distance $z = 10^4$.

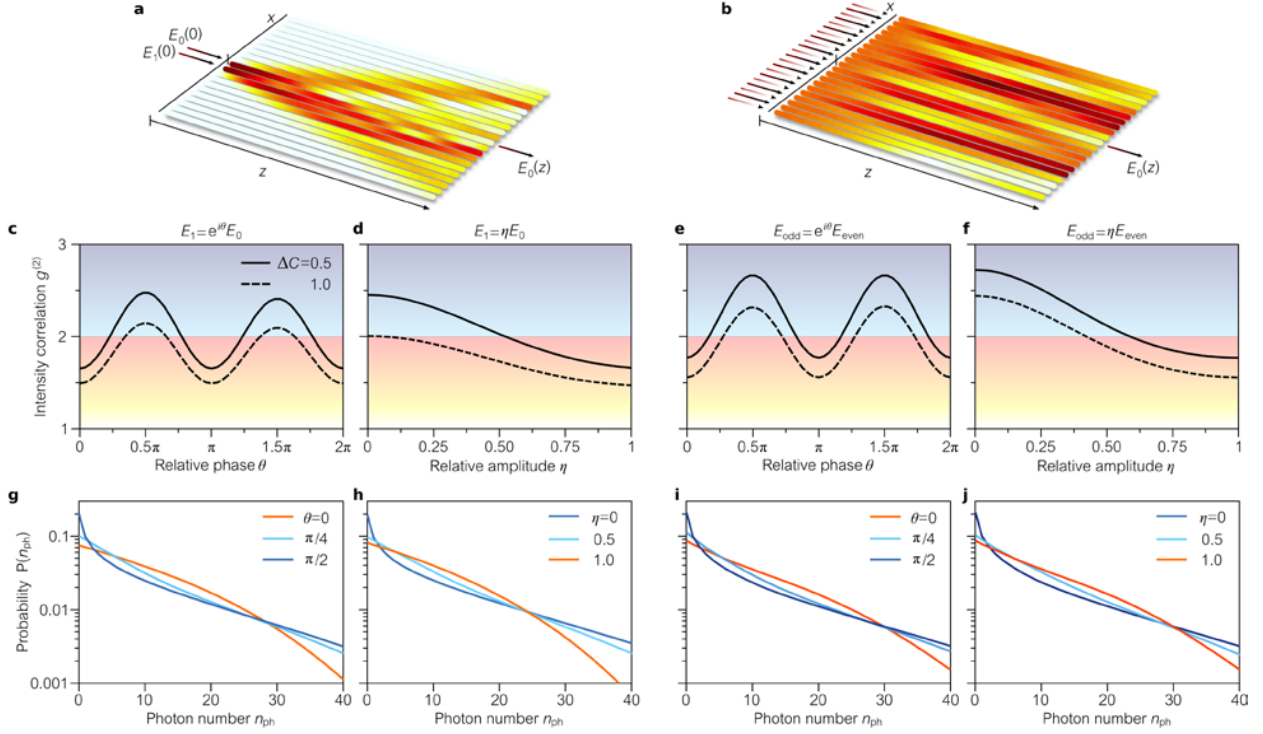

**Figure 4 | Chiral-symmetry-breaking for deterministic control over photon statistics in a disordered lattice. a,b,** Schematic representations of photonic lattices having off-diagonal disorder with (**a**) two-point excitation at $x = 0$ and $1$, denoted $E_0$ and $E_1$, respectively and (**b**) an extended excitation. Color represents the intensity for a single realization in each configuration. **c,d,** Normalized intensity correlation $g_0^{(2)}(z=10)$ as a function of (**c**) the relative phase $\theta$ between $E_0$ and $E_1$ when $|E_0|=|E_1|$ and (**d**) the ratio of the input field amplitudes $\eta = E_1/E_0$ when $E_0$ and $E_1$ are in phase. **e,f,** $g_0^{(2)}(z=10)$ for (**e**) $E_{\text{odd}} = e^{i\theta} E_{\text{even}}$ when $|E_{\text{odd}}|=|E_{\text{even}}|$ and (**f**) $\eta = E_{\text{odd}}/E_{\text{even}}$ when $E_{\text{odd}}$ and $E_{\text{odd}}$ are in phase, where $E_{\text{odd}}$ and $E_{\text{even}}$ denote collectively the odd and even lattice sites, respectively. **g-j,** Photon number distributions corresponding to the tuning of intensity correlations in **c-f** for $\Delta C = 0.5$. **c-j,** Lattice size $2N+1 = 51$ and the ensemble size is $10^6$.